\documentclass[preprint2,flushrt]{aastex}
\usepackage{apjfonts}

\newcommand{\nc}{\newcommand}
\nc{\bec}{\begin{center}}
\nc{\enc}{\end{center}}
\nc{\beq}{\begin{equation}}
\nc{\enq}{\end{equation}}
\nc{\bei}{\begin{itemize}}
\nc{\eni}{\end{itemize}}
\nc{\bee}{\begin{enumerate}}
\nc{\ene}{\end{enumerate}}
\nc{\namely}{{\it viz.}}
\nc{\td}{T$_{\rm d}$}
\def\lsun{$L_\odot${}}
\def\micron{\hbox{$\mu$m}}

\def\dep100{$\tau_{\rm 100}$}
\def\dep150{$\tau_{\rm 150}$}
\def\12CO{$^{12}$CO}
\def\13CO{$^{13}$CO}

\shortauthors{MOOKERJEA et al. }
\shorttitle{FIR observations of Orion A and B}
\slugcomment{Post refereeing revised version dated \today}

\begin{document}


\title{Distribution of Cold dust in Orion A and B }

\author{ B. Mookerjea \altaffilmark{1}, S.K. Ghosh\altaffilmark{1}, T.N. Rengarajan\altaffilmark{1},\\ S.N. Tandon \altaffilmark{2} \& R.P. Verma\altaffilmark{1 } }

\altaffiltext{1}{Tata Institute of Fundamental Research,
            Homi Bhabha Road,
            Mumbai  400 005, India}
\altaffiltext{2}{Inter-University Centre for Astronomy \& Astrophysics,
             Ganeshkhind, Pune 411 007, India}


\begin{abstract}

Large scale  far-infrared (FIR)  observations of the Orion complex  at 205
and 138 \micron\ are presented with an aim of studying the distribution of
cold ($<$25 K) dust.  The maps in these FIR bands extend over $\sim$3600
sq. arcmin and cover regions around OMC-1, 2, 3 in Orion A and NGC 2023 and
NGC 2024 in Orion B. Some limited regions have also been mapped at 57
\micron. A total of 15 sources in Orion A and 14 in Orion B (south) have
been identified from our FIR maps. Dust temperature distribution in both
Orion A and Orion B (south) have been determined reliably using the maps at
205 \& 138 \micron\ obtained from simultaneous observations using almost
identical beams (1\farcm6 dia). These temperatures have been used to
generate map of \dep150, the optical depth at 150 \micron, for the Orion B
region. The coldest source detected is in OMC-3 and has a temperature of
$\sim$ 15 K. The diffuse FIR emission in the different sub-regions is found
to vary between 25 \% to 50 \% of the total FIR emission from that
sub-region.

\end{abstract}




\section{Introduction}

In this paper we present a large scale high angular resolution
($\sim$1\arcmin) study of the continuum emission at 205 and 138 \micron\
from the two Giant Molecular Clouds  Orion A and Orion B (south) in
the Orion complex. The mapped region of Orion A includes OMC-1, 2, 3 and
IRAS 05327-0457 (NGC 1977). In Orion B we have mapped mainly the southern
portion covering NGC 2023 and NGC 2024. Limited regions including, OMC-1
and 2 in Orion A and NGC 2024 in Orion B have also been observed at 57
\micron.

The southern part of the Orion cloud complex, also known as Orion A,
contains star forming cores which are distributed along the north-south
direction (integral shaped ridge) and named OMC-1, 2 and 3 from south
to north. Observations at sub-millimeter/far-infrared wavelengths have
shown that both warm and cold dust with quite different spatial
distribution contribute to the emission from the Orion molecular clouds
\citep{thronson86,chini97}. Cold dust, with temperature $\sim$ 20-25 K,
accounts for bulk of the mass of dust in the Orion clouds and emits
significantly only at wavelengths longer than 100 \micron. Earlier
observations of Orion A concentrated on the emission from the hot dust
viz.,  mapping of OMC-1 at 50\--100 \micron\ by \citet{wern76} and of
M43 at 37\--160 \micron\ by \citet{smith87}. Later observations have
mostly explored the emission from the cold dust in this region. The
most recent of these observations was by \citet{john99} covering OMC-1,
2 and 3 and extending over an area of 50\arcmin$\times$10\arcmin\ with
an angular resolution of $\sim$ 7\farcs5 at 450 \micron\ and 14\arcsec\
at 850 \micron. \citet{lis98} had also mapped a region of
6\arcmin$\times$18\arcmin\ covering OMC-2 and 3  at 350 \micron\ with
an angular resolution of 11\arcsec. However, the pioneering work in
detecting a number of cold sources in OMC-2 and an extremely cold one
($\sim$ 10 K) in OMC-3 was by \citet{chini97}, who  mapped a region of
$\sim$15\arcmin$\times$5\arcmin\ with an angular resolution of $\sim$
11\arcsec\ at 1300 \micron. Most of these condensations are considered
to be protostellar candidates because of the associated outflow
activities \citep{castet95}. Observations in molecular lines like
C$^{18}$O by \citep[][;and references therein]{castet95} have revealed
that the gas in OMC-2 and OMC-3 have kinetic temperatures equal to 24 K
and 19 K respectively, both much smaller than that in OMC-1 (70 K).
These observations also identified both, OMC-2 and OMC-3, to be regions
of high column density. C$^{18}$O observations have also detected
several moderate to strongly collimated bipolar outflows in OMC-3,
which suggests recent star formation activity. The very low
temperatures of sources in both OMC-2 and 3 along with the evidence for
star formation, indicate very early stages of evolution of, probably
intermediate mass stars. In this paper we have explored the
distribution of dust emission and temperature of the Orion A region in
a wavelength regime (205 \& 138 \micron\ in particular) where no high
angular resolution measurements exist in literature.

The Orion B cloud contains  4 major sites of star formation \namely, NGC
2071 \& NGC 2068 in the north and NGC 2023 \& NGC 2024 in the south. We
have observed only the regions around NGC 2024 and 2023. The brightest
emission source in the region is NGC 2024, which is a region of active star
formation. The 1300 \micron\ observations over a region of $\sim$
5\arcmin$\times$5\arcmin\ (with an angular resolution $\sim$ 30\arcsec) by
\citet{mez88} have revealed six condensations in NGC 2024, aligned in the
north-south direction. These condensations also have masers and HII regions
associated with them. The nature of these condensations was an issue of
debate which was resolved by \citet{mez92}. From their observations at 870
\& 1300 \micron\ they concluded that these condensations are indeed
isothermal protostars. \citet{lada91a} presented a survey of Orion B (Lynds
1630) molecular cloud in CS(2-1) line (1\farcm8 angular resolution) and
identified the regions including NGC 2023 and NGC 2024 to be bright in CS,
thus indicating high density of the region. In a follow-up program
\citet{lada91b} surveyed Orion B in the K (2.2 \micron) band and found that
more than 50\% of the NIR sources are associated with the 4 identified star
forming regions. In a recent high angular resolution (22\--30\arcsec)
survey of Orion B (south) at 1300 \micron\ extending over
$\sim$80\arcmin$\times$68\arcmin,  \citet{laun96} have identified some of
the dust emission peaks around NGC 2023 and NGC 2024 with the CS cores
detected by \citet{lada91a}. Of the existing FIR observations of NGC 2023
and NGC 2024, the IRAS observations in the first place trace relatively
warmer dust ($\geq$ 30 K)  and secondly have angular resolutions varying
with wavelength, so that calculating the temperature distribution from the
observed flux densities is not straightforward. On the other hand, the 1300
\micron\ observations, though of high angular resolution, trace extremely
cold dust. This has motivated us to carry out simultaneous high angular
resolution far-infrared observations at 205 and 138 \micron\ with identical
beams so that the density and temperature distribution of dust emitting
significantly in this intermediate regime can be reliably quantified. 

The rest of the paper is organized as follows: Section 2 describes the
observations and Section 3 contains the results and discussion of the same.

\section{Observations}

\subsection{The 205 \& 138 \micron\ maps}

Orion molecular clouds A and B (south) were observed using a two-band
far-infrared photometer system (PHT-12) at the Cassegrain focus of the
TIFR 1-meter (f/8) balloon-borne telescope. The far-infrared telescope
was flown from the TIFR Balloon Facility, Hyderabad, in central India
(Latitude = 17\fdg 47 N , Longitude = 78\fdg 57 E ) in November, 1995.
Details of the telescope and the observational procedure have been given
by \citet{ghosh88}. The photometer (PHT-12) consists of 12 composite
silicon bolometers, each having a FOV of 1\farcm6 and arranged in a
3$\times$2 array for each band. These arrays are cooled to 0.3 K by
liquid $^{3}$He. The sky was chopped at 10 Hz by wobbling the secondary
mirror. The chopper throw was 4\farcm2 along the cross-elevation axis.
The same region of the sky was viewed simultaneously in two bands. The
effective wavelengths for the two bands for a $\lambda ^{-2}$ emissivity
law and a temperature of 35 K are  205 \micron\ and 138 \micron. Saturn
was observed for absolute flux calibration \citep{ghosh88} as well
as determination of the instrumental Point Spread Function (PSF)
including the effect of sky chopping. 

The simultaneous mapping in the two FIR bands was carried out by using
raster scans in which the telescope was continuously moved along the
cross-elevation axis and stepped in elevation. The chopped FIR signals
were gridded in a two dimensional sky matrix (elevation $\times$
cross-elevation) with a cell size of 0\farcm 3$\times$ 0\farcm 3 . The
observed chopped signal matrix was deconvolved using an indigenously
developed procedure based on the Maximum Entropy Method (MEM) similar to
that of \citet{gull78} \citep[see][for details]{ghosh88}. The FWHM sizes
of the deconvolved maps of the point-like source (Saturn) are found to be
1\farcm 6$\times$1\farcm 9 and 1\farcm 6$\times$1\farcm 8 in the 138 and
205 \micron\ bands respectively. An optical photometer was used at the
Cassegrain focal plane to improve the absolute positioning  of the
telescope during the flight. In addition for all the observations
presented here, the signal from this optical photometer measured during
the far-infrared scans have been used during offline analysis to further
improve the aspect information. The absolute positional accuracy achieved
in the FIR maps is $\sim$ 0\farcm5 .

\subsection{The 57 \micron\ map}

Limited regions of both Orion A (OMC-1 \& 2) and Orion B (NGC 2024) were
also observed at 57 \micron\ in an earlier flight (November, 1990) of the
TIFR balloon-borne telescope, using a different photometer (PHT-2). This
two band photometer, PHT-2,  was based on two liquid $^{4}$He cooled
Ge(Ga) bolometers. The FIR passbands were 45--70 \micron\ and 110--160
\micron\ with effective wavelengths 57 \micron\  and 150 \micron\ for an
input source spectrum defined by a greybody of temperature 35 K and
$\lambda^{-2}$ emissivity.  A 2\farcm4 (dia) region of the sky was
simultaneously viewed in both the FIR bands. More details of PHT-2 can be
found in \citet{verma94}.  Here, only the observations at 57 \micron\ are
presented, since the other band is very similar to the shorter waveband of
PHT-12 which has better sensitivity. The PHT-2 was calibrated by observing
Jupiter. Some preliminary results from observations of the Orion region
have been presented by \citet{das93}. The final results are presented
here. The achieved angular resolution in the image processed maps is
$\sim$1\farcm5 as determined from the observations of Jupiter. 

\subsection{The HIRES processed IRAS maps}

To supplement our balloon-borne observations, we have used the IRAS survey
data in all wavebands (12, 25, 60 and 100 \micron) for the regions mapped
by us in Orion B (south). These data were HIRES processed \citep{auman90}
at the Infrared Processing and Analysis Center (IPAC\footnote[1]{IPAC is
funded by NASA as part of  the IRAS extended mission program under
contract to JPL.}, Caltech) for improving the angular resolutions of the
raw maps. 

\section{Results \& Discussion}

\subsection{Analysis of FIR data}

Figure~\ref{bigorib} and Figure~\ref{oria1} respectively show the
regions (indicated by the dotted boundary) in Orion B (south) and Orion
A mapped at 205 and 138 \micron. Regions extending over $\sim$
42\arcmin$\times$38\arcmin\ in Orion A and $\sim$ 38\arcmin
$\times$53\arcmin\ in Orion B (south) have been observed at both these
FIR wavelengths. The 57 \micron\ maps for both sources cover relatively
smaller regions. Reliable dust temperature T(138/205) maps for selected
regions have been generated utilising the simultaneity of observation
and nearly identical beams at 205 and 138 \micron. The dust emissivity
law has been assumed to be  $\epsilon_{\nu}\propto\nu^{\beta}$ and the
method used is briefly described in Appendix 1. For most star forming
regions the emissivity exponent at longer wavelengths ($>$ 100 \micron)
is found to be between 1.5 to 2.5 and the calculated properties of most
common types of dust grains \citep{draine84} show the same at these
wavelengths. Based on these facts we have chosen $\beta$ = 2 for
deriving the T(138/205) distribution. Considering the uncertainties in
the measured flux densities and the wavelength regime probed, the
temperature computed has an uncertainty of $\sim$ $\pm$2K between 14 K
and 50 K and of $\sim$ $\pm$5K between 50 and 70 K. The intensity ratio
of our FIR bands is rather insensitive to temperatures below 10 K and
above 70 K. Above 70 K the estimated temperature has an uncertainty of
$\sim$ $\pm$8K. The morphology of the spatial distribution of \td\ is
insensitive to the assumed value of $\beta$. Only the numerical values
of \td\ depend on $\beta$, e.g., a temperature of 15 K for $\beta$ = 2
corresponds to 17 K for $\beta$ = 1 and \td\ =  30 K for $\beta$ = 2
corresponds to 50 K for $\beta$ = 1. Distribution of \dep150, the
optical depth at 150 \micron\ for Orion B has also been generated using
these dust temperatures and $\beta$ = 2.

We have adopted a well-defined extraction algorithm to identify
discrete sources in these maps. This algorithm is based on
identification of a local maximum followed by a critical study of the
growth curve of the integrated flux densities in consecutive annular
rings with the maximum at the center. The reliability of the sources
detected in each map has been ensured by the following procedure. The
entire observation for each region has been split into two independent
datasets (by considering only the alternate scan lines in each raster)
and processed (deconvolved) separately to generate two maps. The peaks
detected in both these maps within a predetermined positional error
have been used to quantify the dynamic range and also to validate the
detected sources. The final maps presented combine all data and
positions of the ``confirmed sources" in the composite maps have been
determined. These sources have signals more than 10 times the estimated
noise level of the respective maps. In this paper the method of
presenting the sources is as follows : Peaks identified in any two TIFR
wavebands are associated if they lie within 1\arcmin. For association
with IRAS (HIRES) sources the same parameter has been taken to be
1\farcm5. The coordinates quoted for a particular source are for the
position of the peak at the longest (TIFR) wavelength at which it is
detected.

\subsection{Orion A}

Figure~\ref{oria1} shows the 205 and 138 \micron\ intensity distribution
of the mapped region in Orion A. The maps include the Orion Molecular
Clouds 1, 2, 3 and the source IRAS 05327-0457. Figure~\ref{oria58} shows
the intensity map for a limited region in Orion A at 57 \micron.
Table~\ref{oriatab1} presents the coordinates and flux densities (in a
circle of 3\arcmin\ dia) for all the discrete sources detected in the
mapped regions of Orion A at 205 and 138 \micron. In the same table we
also present the longer wavelength associations of these sources along
with the measured flux densities at 1300 \micron\ \citep{chini97}. For the
discrete sources detected at 138, 205 and 1300 \micron\ the dust
temperature (\td) and the emissivity exponent ($\beta$) have been
determined by fitting a single temperature greybody spectrum. \td\ and
$\beta$ for A8 have been determined in a similar manner by using the flux
densities measured at 450, 790 and 1100 \micron\ by \citet{goldsmith97}
along with those from the present work. For all other sources with flux
densities measured only at 205 and 138 \micron\ \td\  has been determined
assuming $\beta$ = 2. For a source not detected in any one of the
wavebands (marked by $\dagger$ in Table~\ref{oriatab1}), we give the flux
density obtained by integrating at the position of the source as detected
in the other waveband(s). Table~\ref{oriatab2} presents the positions,
flux densities and associations of these sources detected at 57 \micron.
The source IRAS 05327-0457 has been studied in detail in the far-infrared
by \citet{bmook00}; here we concentrate mainly on OMC-1, 2 and 3.

Figure~\ref{oriatemp} shows the temperature distribution as calculated
from the observations at 205 and 138 \micron\ for OMC-1 and 2. The
temperature in OMC-1 is as high as $\sim$ 70 K close to the Orion nebula.
The temperature of the OMC-1 and the bar region as calculated by
\citet{wern76} from the ratio of 50 and 100 \micron\ flux densities is
found to be between 55 and 85 K. For the same region, the temperature
calculated from our 205 and 138 \micron\ fluxes is between 45 and 70K.
This is consistent with the fact that the longer wavelength observations
presented here are tracing colder dust component as compared to those
traced  by the 50 and 100 \micron\ observations. The temperature drops to
an average of $\sim$ 25 K in OMC-2 and is as low as 15 K in OMC-3 (not
shown in Figure). The temperature map has been generated by assuming the
dust to be optically thin  but explicit calculation of optical depth (at
150 \micron) very close to the global intensity peak of OMC-1,  shows the
regions to be optically thick. In such cases, the temperatures quoted
would be underestimates by as much as 5K. However, for OMC-1, 2 and 3, the
basic morphology of the isotherms remains unaltered.

Orion A is an ensemble of cores with a variety of temperatures, physical
sizes and masses. This variety in physical conditions makes it conducive
to harbouring stars in different stages of formation and evolution and
hence makes it astrophysically interesting. The far-infrared maps
presented here also support this variety by showing morphological
characteristics strongly dependent on the dust temperature. While the 205
and 138 \micron\ maps tracing colder dust detect FIR emission from OMC-2
and OMC-3 very clearly, the 57 \micron\ map fails to do so. 


OMC-1 (A8 in Table~\ref{oriatab1} and \# 1 in Table \ref{oriatab2}) is the
brightest source detected at all wavelengths. As mentioned earlier, the
submillimeter fluxes at 450, 790 and 1100 \micron\ \citep{goldsmith97}
have been used along with the 57, 138 and 205 \micron\ fluxes to determine
\td\ and $\beta$ for this source. The dust temperature obtained by this
method is 68 K and the $\beta$ is 0.5. If instead we do a similar fitting
omitting the flux density at 57 \micron\ we get \td\ =30 K and $\beta$ =
1.1. Thus the high temperature and low $\beta$ as presented in
Table~\ref{oriatab1} might be due to the effect of attempting to explain
the emission over a wide range of wavelengths using a single temperature
greybody spectrum. In reality there is probably a cold diffuse component
of dust as well which contributes to the emission at longer wavelengths.
In the OMC-1 region the maps at 205 \& 138 \micron\ also detect the source
A5. Although the position of this source at 205 and 138 \micron\ differ by
$\sim$1\farcm5, we have associated the two, since there is no other nearby
source of similar strength. We also note that the temperature of the
source A5 is $\sim$ 80 K, which is even higher than the temperature of A8. 

The peaks \# 2 and \# 3 detected at 57 \micron\ (see
Table~\ref{oriatab2}) are not detected at the longer wavelengths. The
source \# 2 is associated with the HII region M43 and the intensity
distribution around it is similar to that observed at 60 \micron\ by
\citet{smith87}. The source \# 3 is due to the heating of the dust in the
photodissociation region around the trapezium stars. Northern tip of the
emission detected at 57 \micron\ is clearly displaced to the east with
respect to the integral shaped emission ridge detected in OMC-2 and 3. We
can discern local enhancements in T(138/205) map close to the positions
of the peaks \#2 and \#3. These strongly indicate the existence of hotter
dust towards the east.


Of the observations presented here OMC-2 and 3 are detected only at the
longer wavelengths (205 and 138 \micron). In OMC-2 we detect two
sources (A10 and A11) at 138 \micron\ and only one source (A12) at 205
\micron. As suggested by Table~\ref{oriatab1} the sources detected at
our resolution are each associated with more than one submillimeter
(1300 \micron) sources. We note that whereas A10 and A11 are associated
with two mutually exclusive groups of submillimeter sources, A12 is
associated with sources from both groups. The fact that we detect A10
and A11 only at 138 \micron\ and A12 only at 205 \micron\ could be due
to the presence of a diffuse cold component which reduces the contrast
between the sources and the intermediate region. This could effectively
smoothen out the smaller peaks (like A10 and A11) in this region , thus
resulting in  only  one peak (A12) at 205 \micron. The flux densities
at 1300 \micron\ of the detected sub-mm sources have been calculated
using the peak fluxes and size of the individual sources \citep[refer
to Table 1 of ][]{chini97}. For each FIR source the contribution at
1300 \micron\ from all associated sources are combined and presented in
Table~\ref{oriatab1}. For all sources detected in OMC-2 a single
temperature greybody spectral fitting gives 20K$<$\td $<$25K and
1.6$<$$\beta$ $<$ 2.0. Both these are consistent with the results of
\citet{chini97}. Recent radio observations at 3.6 cm by \citet{reip99}
have detected sources coinciding with the peaks in dust emission. These
associations are also presented in Table~\ref{oriatab1}. The radio
emission further substantiates that the sources detected in OMC-2
though reasonably cold are not only regions of high density but also
have YSOs embedded in them.


We have detected one source (A9) in the OMC-3 region. Positionally it
is coincident with the sub-mm source MMS 6 \citep{chini97}, the latter
being the brightest of all the sub-mm sources detected in OMC-3. The
flux density at 1300 \micron\ for the source A9 is estimated in the
same manner as above. A greybody  spectrum fitted  to the
measured flux densities gives \td\ = 15 K  and $\beta$ = 2.0. A9 is thus
the coldest source in Orion A, both \td\ and $\beta$ estimated here are
consistent with \citet{chini97} and \citet{lis98}.

The total far-infrared luminosity has been estimated using the
temperature map (Fig.~\ref{oriatemp}) for the region with OMC-1 and OMC-2
and the intensity maps of the same region. The dust emissivity exponent
assumed for this calculation is 2.0. From the FIR observations presented
here the total luminosity detected in OMC-1 and 2 is
$\sim$3.5$\times$10$^{5}$ \lsun. For a similar region the luminosity based
on 60 and 100 \micron\ observations as given by \citet{stacey93} is
$\sim$5$\times$10$^{5}$ \lsun.  The diffuse emission at these wavelengths
has been calculated by subtracting the contribution of the discrete
sources and is about 25\% of the total luminosity.

The molecular cloud OMC-1 has long been the most extensively studied
region in Orion A. OMC-2 and 3 have been studied only in the recent
past and have been found to be rich and interesting  as a star forming
region. Our observations provide useful constraints on the intensity and
temperature distribution of this region.

\subsection{Orion B (south)}

\subsubsection{Region around NGC 2024}

Figure~\ref{ngc2024} presents intensity maps of the region around NGC 2024
at 205, 138, 100 and 57 \micron. The map at 57 \micron\ is for a smaller
region, but brings out the main features of NGC 2024 at this wavelength.
Table~\ref{oribtab} presents the measured flux densities at the 4 IRAS and
3 TIFR wavelengths for the sources detected in this region. The sub-mm and
IRAS Point Sources associated with these FIR sources, along with the
sub-mm flux densities measured at 1300 \micron\ by \citet{laun96} are
presented in the Table~\ref{oribtab}. In all 7 sources have been
identified with sources in the IRAS Point Source Catalog (PSC) and 2
sources (B3 and B9) have been found  to have sub-millimeter (1300 \micron)
counterparts. For sources not detected in one or two of the TIFR bands
(marked by $\dagger$ in Table~\ref{oribtab}) we have presented the flux
densities obtained by integrating in a circle around the position of the
source as detected at the other wavelength(s). We have fitted a single
temperature greybody spectrum for all sources for which measurement of
flux densities at 3 or more wavelengths longer than 50 \micron\ exist. The
resulting \td\ and $\beta$ are also presented in the same table. For the
remaining sources \td\ is determined using $\beta$ = 2. We find that the
emissivity exponent ($\beta$) for some sources (B6, B10 and B12) are very
small ($\leq$ 0.5). In addition, for  B6 and B12 the fitted \td\ is also
very high. This indicates that for these sources possibly a two or
multi-temperature greybody spctrum is more appropriate.

Observations at 1300 \micron\ by \citet{mez88} have revealed 6
protostellar condensations within  an area of 3\arcmin\ $\times$ 3\arcmin\
around NGC 2024. The far-infrared maps presented here being of limited
angular resolution do not resolve these condensations. However the 205
\micron\ and also the 100 \micron\ maps show evidence of north-south
extension with the northern tip having a slight westward bend
\citep[position of FIR-1, 2 ;][]{mez88}. Positionally B9 coincides with
FIR-5 \citep[brightest condensation detected by][]{mez88} and also with
LBS 33SM, a CS core detected by \citet{lada91a} (hereafter LBS), later
detected at 1300 \micron\ by \citet{laun96}. The HIRES map at 100 \micron\
shows possible signs of anomaly, particularly near the peak. The flux
density measured (in 3\arcmin\ dia) from the 100 \micron\ map shows a very
low value of $\sim$18 kJy  compared to both the IRAS PSC value (35.3 kJy)
and results from the KAO observations by \citet{thron84} (85 kJy). We have
used the measurements of \citet{thron84} in the Table~\ref{oribtab} as
well as for determining \td\ and $\beta$ for B9.

Figure~\ref{ngc2024temp} shows the dust temperature distribution for this
region. It has been generated using the same method as for Orion A. It
shows several isolated peaks with temperatures varying between 60 to 70 K
for most of them. The source B3 is quite cold with a maximum of 30 K and
	going below 15 K towards the edges. The \dep150\ map
	(Figure~\ref{ngc2024temp}) shows very interesting structure close
	to the main source B9. It distinctly shows that the region around
	B9 has an elongated structure with indications of at least 2
	condensations. This structure is very similar to the structure
	shown by the NH$_{\rm 3}$ column density map generated by
	\citet{schulz91}. This very well substantiates the fact that these
	are indeed protostellar condensations since they are so luminous in
	the infrared as well. The cold source B3 is found to be mostly
	optically thin (peak 0.07) at 150 \micron.

The total luminosity of the region calculated using the same method as for
Orion A (Section 3.1) is 1.0$\times$10$^{5}$ \lsun. The diffuse emission
contributes about 50\% (at both the wavelengths) of the total luminosity.
This is substantially larger than the value measured in Orion A (Sec 3.1)
and in other young Galactic star forming regions \citep[$\sim$ 35\% in
W31;][]{ghosh89} . The morphology of the [CII] (158 \micron) emitting
region in Orion B (south) as observed by \citet{jaffe94} agrees very well
with the maps presented in this paper. The total [CII] luminosity detected
for their assumed distance of 415 pc is 230\lsun, while we have  estimated
	the total far-infrared luminosity for a distance of 450 pc.
	Correcting for this difference in distances assumed we find the
	ratio of the [CII] luminosity to the total FIR luminosity to be
	3$\times$10$^{-3}$.

The observations presented here, being on a large scale and of good
angular resolution would be useful as guidelines in understanding the star
formation scenario vis-a-vis the temperature distribution for future large
scale sub-millimeter continuum observations.

\subsubsection{Region around NGC 2023}

Figure~\ref{ngc2023} shows intensity maps of the region surrounding NGC
2023 at 205 and 138 \micron\ alongwith HIRES maps at 100 and 60 \micron.
The maps at 205 and 138 \micron\ are generated by masking NGC 2024 prior to
MEM deconvolution of  the larger map so that details of the fainter source
NGC 2023 are better visible and comparison with 100 and 60 \micron\ HIRES
maps is easier. The source NGC 2023 appears to have a double structure at
205 \micron. It is of interest to note that the map of fluorescent H$_{\rm
2}$ observed by \citet{gatley87} exhibits a shell structure with main peaks
separated by similar extent as B13 and B14. The intensity map at 138
\micron\ however shows only a single source close to the brighter source
(B14) at 205 \micron. Diffuse emission extending southward from the main
source is clearly visible at 60, 100 and 205 \micron, while at 138 \micron\
the limited dynamic range of the map restricts this emission to a seemingly
isolated peak. Table~\ref{oribtab} presents the  positions, flux densities
and associated sources for the 2 peaks (B13 and B14) detected in this
region. The coordinates of the global peaks in the 60 and 100 \micron\ maps
is found to be $\sim$ 5$^{\rm h}$ 39$^{\rm m}$ 7 \arcsec  - 2\arcdeg\
17\arcmin\ 44\arcsec. These global peaks are separated from B14 by more
than 1\farcm5. Since there is no other detected source of such brightness
in the neighbourhood and observations at 100 and 138 \micron\ trace dust of
similar temperature we have associated B14 with NGC 2023. Both sources B13
and B14 have been detected at 1300 \micron\ by \citet{laun96}, but there is
no IRAS detection of B13. For both sources we have used the available flux
densities at wavelengths longer than 60 \micron\ to estimate \td\ and
$\beta$ from a single temperature greybody spectrum (Table~\ref{oribtab}).
The flatness of $\beta$ for B14 indicates the  existence of dust components
of at least two temperature ranges. The source B13 is found to be
reasonably cold (\td\ = 16 K), thus justifying it's non-detection in the
IRAS wavebands.

Figure~\ref{ngc2023temp} shows the T(138/205)  and \dep150\ maps for the
region around NGC 2023 (derived assuming $\beta$ = 2). The T(138/205) map
shows a peak which does not coincide with the intensity peaks at 205 and
138 \micron. The T(138/205) peak, however coincides with the peak positions
at 60 and 100 \micron. The temperature of most of the region is around 30 K
and in places goes down to values as low as $\sim$ 20 K. The \dep150\ map
more or less traces the high intensity regions. The \dep150\ displayed for
regions which lie outside the contours of T(138/205) map are for an assumed
temperature of 10 K (not plotted in the T(138/205) map).

NGC 2023,  being a well known reflection nebula with a photodissociation
region (excited by HD 37903) has been studied in various molecular and fine
structure ([CII] at 158 \micron) lines. The continuum maps presented in
Figure~\ref{ngc2023} are morphologically similar to the C$^{\rm 18}$O map
by \citet{wyr97} and CO (J=2\--1) map by \citet{white90}. The [CII]
observations at 158 \micron\ by \citet{howe91} and \citet{jaffe94} detect
emission which is approximately 25 \% of the emission from NGC 2024. The
rise in temperature, T(138/205), to the north-west of the main peak is
explained by the presence of the exciting star HD 37903. The total
luminosity measured in this region is 1.5$\times$10$^{3}$ \lsun. The
diffuse emission contributes about 35\% of the total emission, which is
smaller than the contribution of diffuse emission around NGC 2024. The
total [CII] luminosity from the region as detected by \citet{howe91} is 7
\lsun\  and is equal to $\sim$ 0.5\% of the total FIR luminosity estimated
here. This is consistent with results from most of the Galactic star
forming regions as given by \citet{howe91}. Detection of low temperature in
the region around NGC 2023 alongwith emission in the far-infrared,  support
ongoing formation of low to intermediate stars in the region. Indication of
more than 50 \% of the emission being in the condensations further
indicates star formation in clumps in this region.

\section{Summary}

In this paper we have presented large scale high angular resolution
($\sim$1\arcmin) far-infrared (205 and 138 \micron) maps of the Orion A
and Orion B (south) molecular clouds. For limited regions, maps at 57
\micron\ have also been presented. HIRES processed IRAS images at 100 and
60 \micron\ of the Orion B have been used for comparison. The regions
covered include OMC-1, 2, 3 in Orion A and NGC 2024 and 2023 in Orion B
(south). In all  15 condensations in Orion A and 14 condensations in Orion
B (south) have been identified. In OMC-3 the very cold source MMS 6
\citep{chini97} was detected at both 205 and 138 \micron\ and the
temperature was determined to be $\sim$15$\pm$2 K. The far-infrared
counterparts of the extremely dense CS cores \citep{lada91a} LBS 33, LBS
34, LBS 36 and LBS 40 were detected in Orion B (south). The total
far-infrared luminosity and the relative contribution of diffuse emission
to the luminosity of these regions have also been quantified.

\acknowledgements

We thank B. Das for his help during initial stages of analysis of the 57
\micron\ data and S.L. D'Costa, M.V. Naik, D.M. Patkar, M.B. Naik, S.A.
Chalke, G.S. Meshram \& C.B. Bakalkar for their support for the
experiment. The members of TIFR Balloon Facility (Balloon group and
Control \& Instrumentation group), Hyderabad, are thanked for their roles
in conducting the balloon flights. Thanks are due to IPAC for providing
HIRES processed IRAS data. We thank the anonymous referee for the
suggestions which have improved the quality of the paper.

\appendix

\section{Estimation of dust temperature distribution using the two band FIR
observations}

As mentioned in Section 2.1 the spatial intensity distribution in each
band is gridded with pixels of size 0\farcm3$\times$0\farcm3. Only those
pixels which have signals exceeding 5 times the noise level of the
corresponding map are considered to be ``valid" for the purpose of finding
the dust temperature distribution. Each intensity map is first smoothed by
taking a running average over 3$\times$3 ``valid" pixels. This is done to
be extremely conservative about the derived spatial structures in the
temperature maps. Since the two intensity maps refer to identical portions
of the sky and the beams are also almost identical, a ``pixel by pixel"
ratio between the two maps gives reliable dust temperatures.

For optically thin emission the observed flux density at any wavelength can
be written as

\beq
F_{\nu} = \Omega B_{\nu}(T_{d}) \tau_{\nu}
\enq

where $\Omega$ is the solid angle of the region under consideration. 

We assume $\tau_{\nu}$ $\propto$ $\nu^{\beta}$. Since the pixel sizes are
identical the ratio of flux densities at any two wavelengths would be:

\beq
\frac{F_{\nu_{1}}}{F_{\nu_{2}}} = \left (\frac{\nu_{1}}{\nu_{2}}\right )^{3+\beta}
\frac{[exp(h\nu_{2}/kT_{d}) -1]}{[exp(h\nu_{1}/kT_{d}) -1]}
\enq

Thus for an assumed emissivity exponent ($\beta$) the ratio of the flux
densities at any two wavelengths is a function of the dust temperature
only. The temperature is computed using a lookup table containing the
calculated ratios of flux densities for various temperatures and an
assumed $\beta$. This process is repeated for all  pixels to generate
the dust temperature distribution map.

%


\newpage

\begin{center}
{\bf \large Figure Captions}
\end{center}

\figcaption[fig1a.ps]
{Intensity map of Orion B (South) at 205 \micron. Dotted boundary shows the
region mapped. The peak is 2.50 kJy/sq. arcmin. The contour levels are at
90 to 30\% (in steps of 20), and 20, 10, 5, 2.5, 1.25 and 0.63\% of the
peak.
\label{bigorib}
}

\figcaption[fig2.ps] 
{The intensity maps of Orion A. Peaks are 9.00 kJy/sq. arcmin at 205
\micron\ and  29.1 kJy/sq. arcmin at 138 \micron. The contour levels are
same as in Figure~\ref{bigorib}, with an additional contour at 0.31\% of
the peak. The lowest contour level is 3 times the measured noise.
\label{oria1} }

\figcaption[fig3.ps] 
{The intensity map of the Orion A at 57 \micron. The peak is
59.6 kJy/sq. arcmin The contour levels are same as in
Figure~\ref{oria1}.
\label{oria58} }

\figcaption[fig4.ps] {The dust temperature map for the region covering
OMC-1 and 2, assuming a dust emissivity of $\epsilon_{\nu}\propto\nu^{2}$.
The lowest temperature found in this region is 14 K. The numbers written
on the contours refer to the temperatures, with the highest contour
displayed being at 70 K.
\label{oriatemp} }

\figcaption[fig5.ps]
{The intensity maps of the region around NGC 2024. Peaks are (a) 2.50 kJy/sq.
arcmin at 205\micron, (b) 7.53 kJy/sq. arcmin at 138 \micron, (c) 3.80 kJy/sq.
arcmin  at 100 \micron\ and (d) 19.1  kJy/sq. arcmin at 57 \micron\ The
contour levels (a) and (b) are same as in Figure~\ref{bigorib}. Levels in (c)
are 60 to 10 \% (in steps of 10)  and 5, 2.5 and 1\% of its peak. Levels in
(d)  are 90, 75, 60, 40, 20, 10, 7.5, 5, 3.5, 2, 1 and 0.7\% of the peak.
\label{ngc2024} }

\figcaption[fig6.ps] 
{(a) Same as in Fig~\ref{oriatemp}, except being for the region around NGC
2024. (b)  \dep150\ map for the same region. Contour levels are
 90 to 10 \% (in steps of 10) and 5, 2.5 and 1\% of its peak (0.24).
\label{ngc2024temp} }

\figcaption[fig7.ps] 
{The intensity maps for the region around NGC 2023. The peaks are (a) 325
Jy/sq. arcmin at 205 \micron, (b) 329 Jy/sq. arcmin at 138 \micron, (c)
511 Jy/sq. arcmin at 100 \micron\ and (d) 786 Jy/sq. arcmin at 60 \micron.
The contour levels in (a) are 95\%, 90 to 10\% (in steps of 10) and 5\% of
the peak and in (b) are 95\%, 90 to 30 \% (in steps of 10) of the peak.
The contour levels in (c) and (d) are the same as in Fig. 6b except for an
additional contour at 95 \% of the peak. 
\label{ngc2023} }

\figcaption[fig8.ps]
{(a) T(138/205) map for the region around NGC 2023. The highest contour
displayed is at 40 K. (b) \dep150\ map for NGC 2023 with peak value of
0.14. Contour levels are same as in Fig. 7c.
\label{ngc2023temp}
}



\newpage
\begin{deluxetable}{ccclrrrrr}
\tablewidth{0pt}
\tablecaption{Position, flux densities (in 3\arcmin\ dia), associations, \td\ \& $\beta$ of 
sources identified in Orion A at 205 and 138 \micron\ 
\label{oriatab1}}
\tablehead{
\multicolumn{1}{c}{Source \#} &
\multicolumn{2}{c}{Position\tablenotemark{a}~~(1950)}&
\multicolumn{1}{c}{Association\tablenotemark{b}} &
\multicolumn{1}{r}{F$_{138}$ }&
\multicolumn{1}{r}{F$_{205}$ } &
\multicolumn{1}{r}{F$_{1300}$ } &
\multicolumn{1}{r}{T$_{\rm d}$\tablenotemark{c}} &
\multicolumn{1}{r}{$\beta$\tablenotemark{c}} \\
\colhead{} &
\colhead {R.A.} &
\colhead {Dec.} &
\colhead {}&
\colhead {~~kJy} &
\colhead {~~kJy} &
\colhead {~~~Jy} &
\colhead {~~K}
}
\startdata
A1& 5 31 51.6&  - 5 06 22   & IRAS 05318-0506 & 1.26 & 0.58 & \nodata  & 29 & \nodata\\
A2& 5 32 02.0&  - 4 56 30   & \nodata &  1.24$^{\dagger}$ &0.96 & \nodata  & 20& \nodata\\
A3& 5 32 04.6&  - 5 19 14   & \nodata & 0.83 &  0.39 & \nodata  & 28& \nodata\\
A4& 5 32 11.6&  - 5 02 28   & \nodata & 0.56 &  0.36 &\nodata  & 22 & \nodata\\
A5& 5 32 32.8&  - 5 22 57   & \nodata & 2.65 & 0.68 & \nodata   & 80& \nodata\\
A6& 5 32 43.7&  - 4 57 42   & IRAS 05327-0457 &  2.12 & 0.96 & \nodata & 30& \nodata\\
A7& 5 32 47.3&  - 4 58 33   &\nodata  & 2.06 & 0.86$^{\dagger}$&  \nodata  & $>$32& \nodata\\
A8& 5 32 47.7&  - 5 24 51   & OMC-1  & 112.50 & 35.70 & \nodata  & 68\tablenotemark{d}& 0.5\tablenotemark{d}\\
A9& 5 32 55.6&  - 5 04 00   &  OMC-3,MMS 4,5,6,VLA 3 & 0.42 &  0.53 & 4.77 & 15 & 2.0 \\
A10& 5 32 59.5&  - 5 14 25   & FIR 6a,6b,6c,6d,VLA 14& 2.22 & 1.44$^{\dagger}$ &7.45 &  25 & 1.6\\
A11& 5 33 01.4&  - 5 11 16   &FIR 2,3,4, VLA 11,12 & 1.88 & 1.50$^{\dagger}$& 6.94  &  20 & 1.9\\
A12& 5 33 01.6&  - 5 13 01   &FIR 4,5,6a,6b &  1.98$^{\dagger}$ & 1.71 & 10.4 &  20 & 1.8\\
A13& 5 33 07.1&  - 5 01 49   & \nodata & 0.43$^{\dagger}$ & 0.54 & \nodata & 15 &\nodata \\
\enddata
\tablenotetext{\dagger}{Not detected as a local peak in this waveband. Flux density is
obtained by integrating in a circle of 3\arcmin\ dia around the coordinates tabulated.}
\tablenotetext{a}{Units of right ascension are hours, minutes and seconds, and 
units of declination are degrees, arcminutes, and arcseconds.}
\tablenotetext{b}{IRAS : IRAS PSC ; FIR \& MMS : \citet{chini97}; 
VLA : \citet{reip99}.}
\tablenotetext{c}{Single temperature greybody spectrum; if flux densities
at only 2 bands are available emissivity $\epsilon_{\nu}$ $\propto$
$\nu^{2}$ is assumed. }
\tablenotetext{d}{Using F$_{57}$, F$_{138}$ \& F$_{205}$ (this work) and
F$_{450}$ = 19.8 kJy, F$_{790}$ = 2.671 kJy \& F$_{1100}$ = 0.989 kJy
\citep{goldsmith97}.}

\end{deluxetable}


\begin{deluxetable}{ccccc}
\tablewidth{0pt}
\tablecaption{Position and flux densities (in 3\arcmin\ dia) of sources identified in Orion A at 57 \micron\ 
\label{oriatab2}}
\tablehead{
\multicolumn{1}{c}{Source \#} &
\multicolumn{2}{c}{Position\tablenotemark{a}~~(1950)}&
\multicolumn{1}{r}{F$_{57}$ }&
\multicolumn{1}{c}{Associations}\\
\colhead{} &
\colhead {R.A. } &
\colhead {Dec. }&
\colhead {kJy}&
\colhead {} 
}
\startdata

 1& 5 32 47.3&  - 5 24 27   & 219 & OMC-1 \\
 2& 5 33 00.2&  - 5 16 18 &  5.91 & M43\\
 3& 5 33 03.3&  - 5 21 46   & 3.99 & \nodata\\
\enddata
\tablenotetext{a}{Units of right ascension are hours, minutes and seconds, and 
units of declination are degrees, arcminutes, and arcseconds.}
\end{deluxetable}

\newpage


\begin{deluxetable}{ccccrrrrrrrrrr}
\tablewidth{0pt}
\tabletypesize{\footnotesize}
\tablecaption{Position, flux densities (in 3\arcmin\ dia), associations, \td\ \& $\beta$ of 
sources identified in Orion B at 205 and 138 \micron\ 
\label{oribtab}}
\tablehead{
\multicolumn{1}{l}{Source \#} &
\multicolumn{2}{c}{Position\tablenotemark{a}~~(1950)}&
\multicolumn{1}{c}{Association\tablenotemark{b}}&
\multicolumn{1}{r}{F$_{12}$ }&
\multicolumn{1}{r}{F$_{25}$ }&
\multicolumn{1}{r}{F$_{57}$ }&
\multicolumn{1}{r}{F$_{60}$ }&
\multicolumn{1}{r}{F$_{100}$ }&
\multicolumn{1}{r}{F$_{138}$ }&
\multicolumn{1}{r}{F$_{205}$ }&
\multicolumn{1}{r}{F$_{1300}$ }&
\multicolumn{1}{r}{T$_{\rm d}$ \tablenotemark{c} } &
\multicolumn{1}{c}{$\beta$\tablenotemark{c}} \\
\colhead { } &
\colhead {R.A. } &
\colhead {Dec. }&
\colhead {} &
\colhead {~~Jy} &
\colhead {~~Jy} &
\colhead {~~kJy} &
\colhead {~~kJy} &
\colhead {~~kJy} &
\colhead {~~kJy}&
\colhead {~~kJy}&
\colhead {~~Jy}&
\colhead {}&
\colhead {}
}
\startdata
NGC 2024 &&&&&&&&&&&&\\
region &&&&&&&&&&&&\\
 B1& 5 38 21.4 & - 1 50 19  & 05383-0150 & 60 & 65 & NM\tablenotemark{d} & 0.95 & 1.40 & 0.67& 0.29 & \nodata & 32 & 2.5  \\
 B2& 5 38 21.9 & - 1 54 24 &\ldots  & 68 & 68 & NM & 1.17 & \ldots & 1.04 & 0.30 &\nodata &  25 & 3.8 \\
 B3& 5 38 43.0 & - 1 49 55 &05387-0149 & 80 & 133 & \ldots & \ldots & \ldots & 1.69 & 1.13  & 2.6 &  20 & 2.3 \\
&&&LBS 40 SM  &&&&&&&&&\\
 B4& 5 38 44.5 & - 1 58 44 & \ldots & 87 & 120 & 4.50$^{\dagger}$ & \ldots & 2.79 & 2.18 & 0.67 &\nodata &  21 & 3.8 \\
 B5& 5 38 52.6 & - 1 46 10 &05388-0147 & 89 & \ldots & NM & \ldots & \ldots & 0.57$^{\dagger}$& 0.49 &\nodata & 20  & \nodata\\
 B6& 5 38 57.2 & - 1 54 00 & \ldots & 110 &\ldots &5.51 &\ldots &\ldots &1.89 & 0.90 &\nodata &  95 & 0.4 \\
 B7& 5 38 59.4 & - 1 47 52 & \ldots & 81 & 175 & NM & 2.83 & 3.59 & 1.25 & 0.37$^{\dagger}$ &\nodata &  27  & 3.9\\
 B8& 5 39 08.4 & - 1 52 13 & 05391-0152 & 248 & 1060 & 6.62 & 10.74 &\ldots &  4.24 & 1.31 &\nodata &  32 & 2.8\\
 B9& 5 39 13.7 & - 1 56 52 & 05393-0156 & 86 & 916 & 71.70 & \ldots & 85.00\tablenotemark{e} & 32.30 & 12.06  & 115 & 59 & 1.0\\
&&&LBS 33 SM &&&&&&&&&\\
B10& 5 39 21.8 & - 1 45 47 & 05392-0145 & 17 & 31 & NM & 0.32 & \ldots & 0.83 & 0.59  & \nodata & 35 & 0.5\\
B11& 5 39 21.8 & - 1 49 13  &  05393-0150 & 80 & 176 & 0.71$^{\dagger}$ & 2.10 & 2.96 & 1.56 & 0.65  &\nodata & 32 & 2.4\\
B12& 5 39 28.5 & - 2 01 36  & \ldots & \ldots & \ldots &\ldots & \ldots & 0.77 & 0.51 & 0.27$^{\dagger}$& \nodata & 80 & 0.2\\
\tableline
NGC 2023 &&&&&&&&&&&&\\
region &&&&&&&&&&&&\\
B13& 5 39 02.9 & - 2 20 16 & LBS 36 SM & 25 & 52 & NM &\ldots &\ldots &0.79$^{\dagger}$ & 0.82 & 4.7 & 16 & 2.1\\
B14& 5 39 14.6 & - 2 19 08 & LBS 34 SM & 14 & \ldots & NM &\ldots & \ldots & 1.93 & 1.41 & 12.5 & 25 & 1.3\\
\enddata
\tablenotetext{\dagger}{Not detected as a local peak in this waveband. Flux density is
obtained by integrating in a circle of 3\arcmin\ dia around the coordinates tabulated.}
\tablenotetext{a}{Units of right ascension are hours, minutes and seconds, and 
units of declination are degrees, arcminutes, and arcseconds.}
\tablenotetext{b}{LBS SM : CS cores \citep{lada91a} also detected at 1300 \micron\ by
\citet{laun96}. IRAS PSC for all other entries.}
\tablenotetext{c}{Single temperature greybody spectrum; if fluxes at only 2 bands are
available emissivity $\epsilon_{\nu}$ $\propto$ $\nu^{2}$ is assumed. }
\tablenotetext{d}{Not mapped at 57 \micron. }
\tablenotetext{e}{\citet{thron84}.}  
\end{deluxetable}

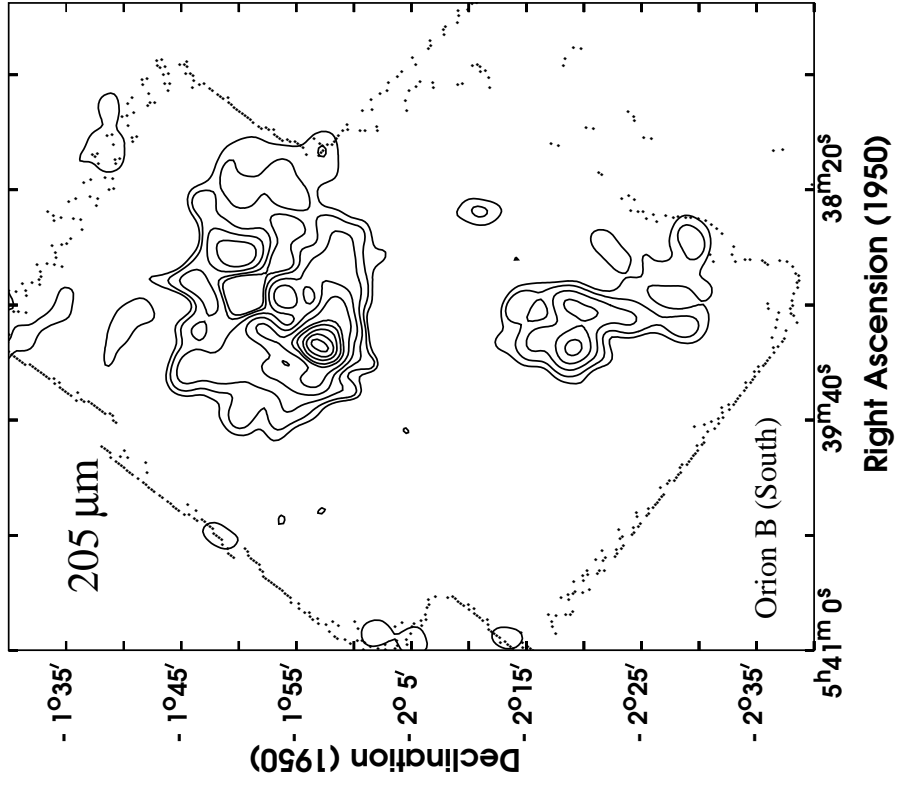
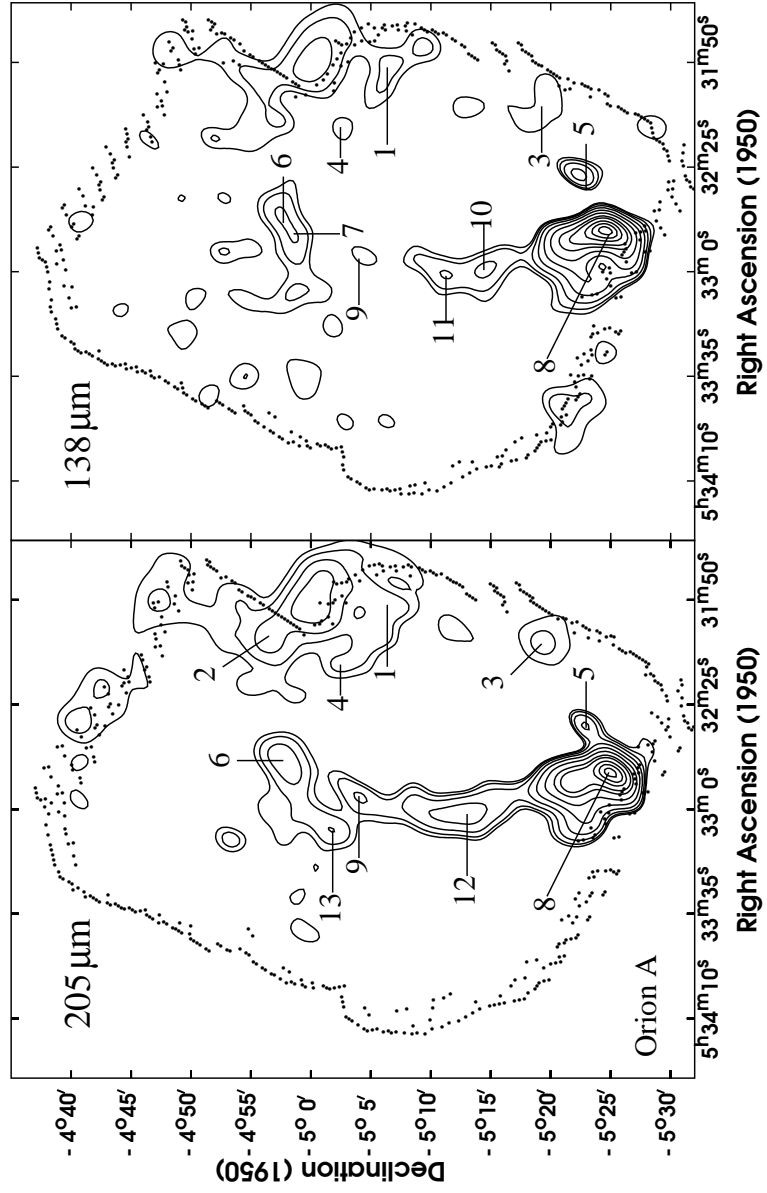
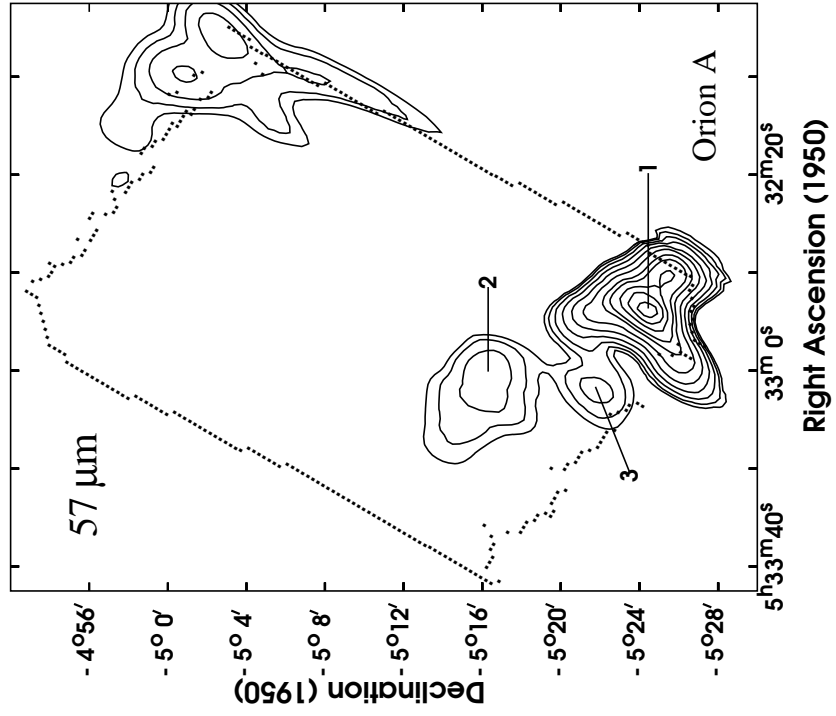
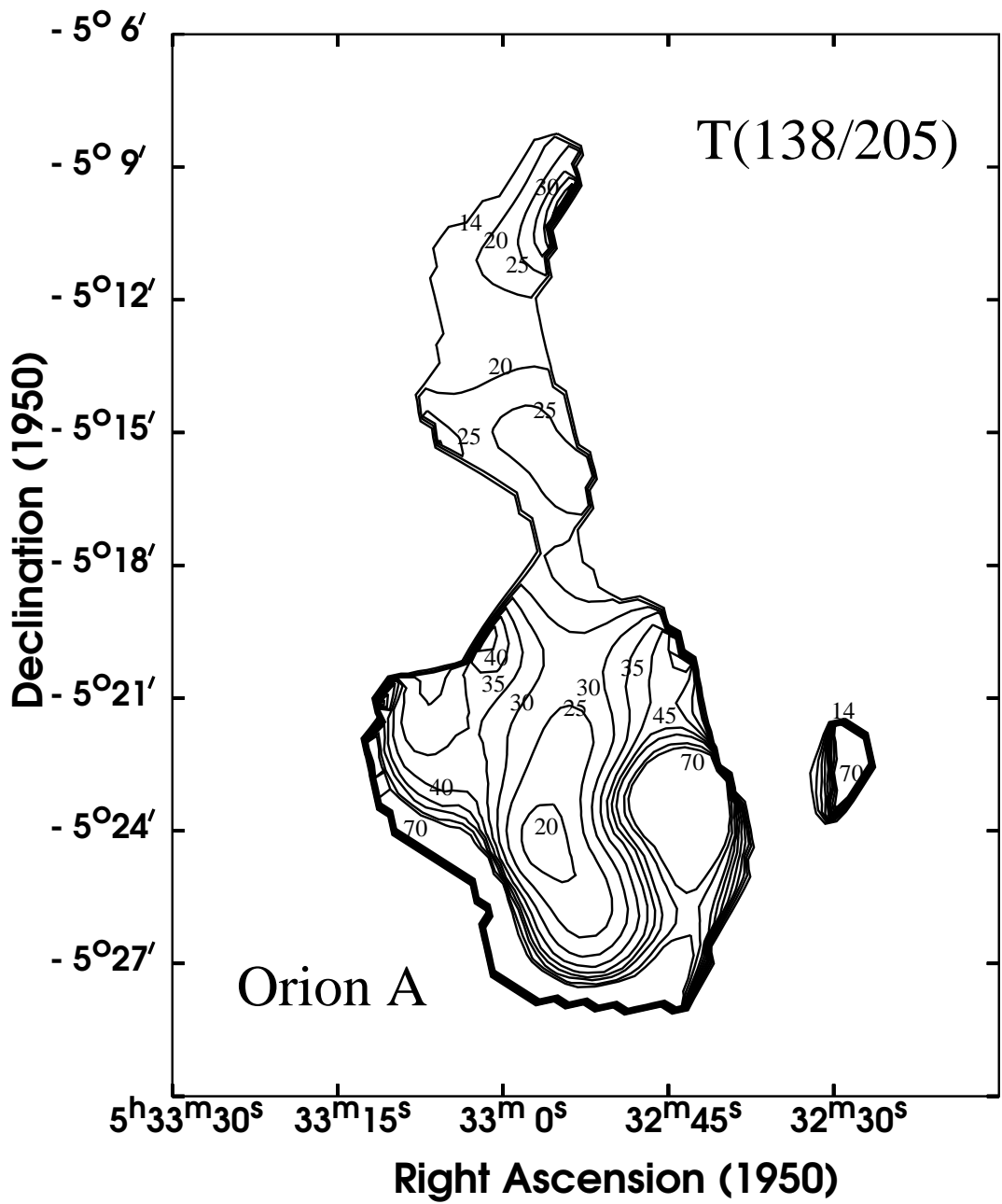
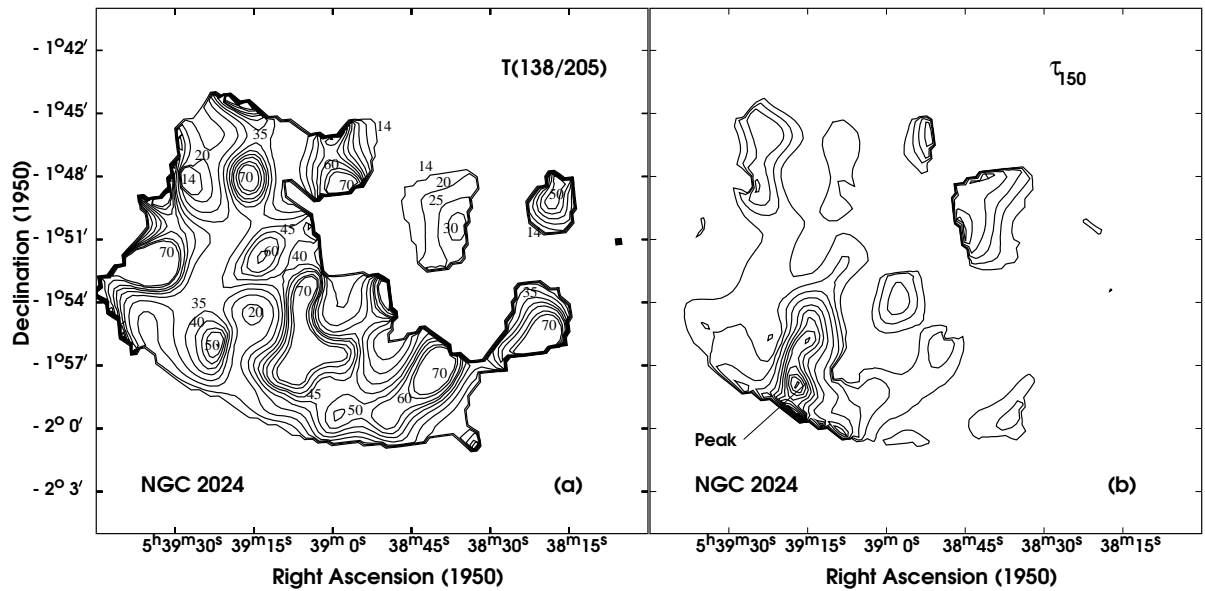
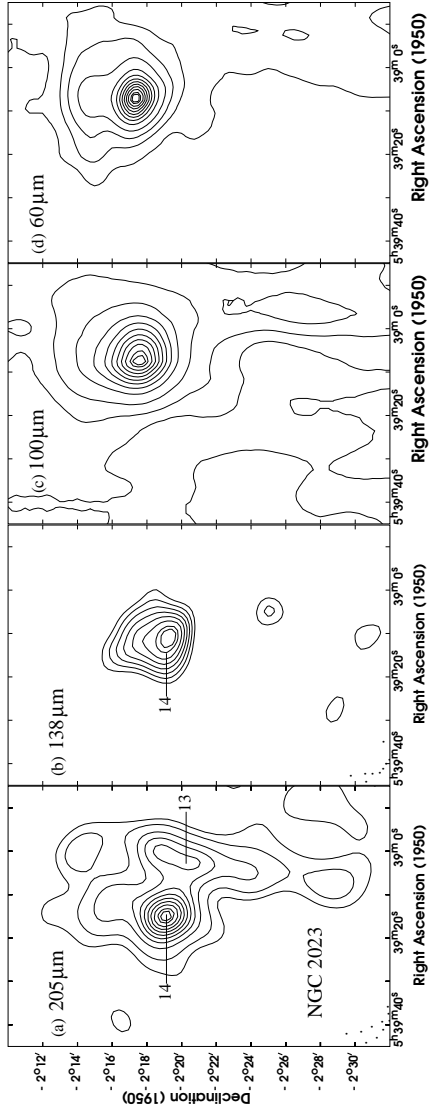
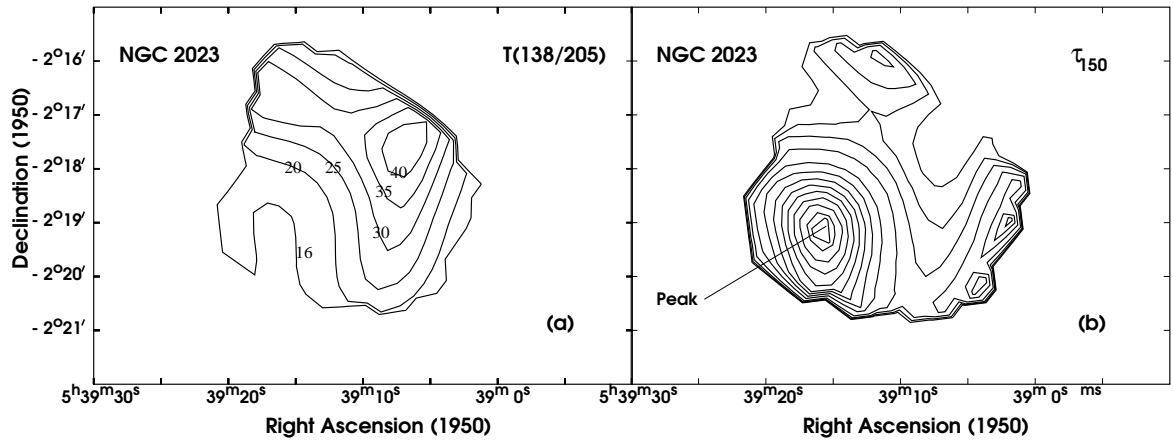
\end{document}